\documentstyle[prl,aps]{revtex}

\draft
\tighten
\twocolumn

\begin{document}

\date{\today}

\title{Implicit Purification for Temperature-Dependent Density Matrices}

\author{Anders~M.~N.~Niklasson}

\address{
Theoretical Division, Los Alamos National Laboratory,
Los Alamos, NM 87545, USA}

\maketitle

\begin{abstract}
{\bf An implicit purification scheme is proposed for calculation
of the temperature-dependent, grand canonical single-particle 
density matrix, given as a Fermi-Dirac operator expansion 
in terms of the Hamiltonian.
The computational complexity is shown to scale 
with the logarithm of the polynomial order of the expansion, or 
equivalently, with the logarithm of the inverse temperature. 
The system of linear equations that arise in each implicit purification
iteration is solved efficiently by a conjugate gradient solver.
The scheme is particularly useful in connection with linear scaling
electronic structure theory based on sparse matrix algebra. 
The efficiency of the implicit temperature expansion technique 
is analyzed and compared to some explicit purification methods 
for the zero temperature density matrix.}
\end{abstract}
~\\


Linear scaling electronic structure theory in combination
with tight-binding, self-consistent Hartree-Fock or density 
functional theory has become a very powerful tool for studying 
complex large material systems \cite{Goedecker99,Wu02}. There are
several ways to achieve a computational cost that
scales linearly with system size. Here we focus
on methods based on the single-particle density matrix for
band-gap materials, the matrix elements of which decay
exponentially with overlap distance. For large systems 
the number of matrix elements above some numerical threshold
therefore scales as ${\cal O}(N)$.
In these schemes the two major steps are the construction
of the tight-binding, Fockian or Kohn-Sham Hamiltonian $H({\bf r,r'})$ 
and the calculation of the density matrix $\rho({\bf r,r'})$. 
The present article concerns aspects of the second 
problem, the construction of the density matrix.

The relation between the density matrix at $T=0$ and
the Hamiltonian is given by the Heaviside step function
\cite{Notation}
\begin{equation}
\rho = \theta(\mu I-H).
\end{equation}
In density matrix schemes this relation is approximated
by constructing $\rho$ from $H$ using sparse matrix algebra, 
where each major operation
computationally scales linearly with the system size, thanks to 
${\cal O}(N)$ matrix sparsity. This can be achieved 
through constrained minimization schemes \cite{Li93,Kohn96},
spectral projections or purification methods
\cite{McWeeny60,Clinton69,Palser98,Holas01,Niklasson02,Niklasson03,NiklassonSP4}, or by
an expansion of the temperature dependent Fermi-Dirac 
function or similar step function  approximations
\cite{Goedecker94,Silver94,Wang94,Silver96,Baer97,HG_ACS}. 
Contour integral representations of the Fermi distribution with 
complex Pad\'{e} polynomials as resolvents, that can be calculated 
${\cal O}(N)$ implicitly \cite{Goedecker93}, as well as combinations 
of various approaches have also been explored 
\cite{Challa99,Bowler99,Helgaker00,Head_Gordon03,Shao03,Daniels99}.

The quadratically convergent purification techniques 
have turned out to be some of the most efficient approaches
for the construction of the density matrix, 
both in memory and speed \cite{Palser98,Daniels99,Niklasson03,NiklassonSP4},
with a computational complexity, in terms of number of 
matrix-matrix multiplications necessary to reach convergence, 
that scales linearly with the logarithm of the inverse band gap 
and the degree of expansion, and with a numerical
error that scales linearly with the threshold \cite{Niklasson03,NiklassonSP4}. 
The Fermi-Dirac operator expansion-based methods based on Chebychev
expansion techniques are generally much slower, with the
computational cost scaling at best with the square root of the degree
of expansion \cite{Paterson73,HG_ACS}. 
However, these methods have an important advantage;
they can account for a finite temperature distribution of the
density matrix. Here we propose an expansion scheme
that combines the low logarithmic complexity and quadratic
convergence of the purification schemes with the finite
temperature Fermi-Dirac distribution. We show how this
can be accomplished by an implicit purification scheme,
based on a Pad\'{e} approximation of the rescaled Fermi-Dirac function,
with a computational complexity that scales logarithmically 
with the expansion order, or equivalently, the inverse temperature.


The Fermi-Dirac distribution \cite{Notation},
\begin{equation}
\Phi_{\rm FD}(\varepsilon,\mu,\beta) = \frac{1}{e^{\beta (\varepsilon - \mu)} + 1},
\end{equation}
occurs in statistical mechanics as the occupational distribution of 
fermions at finite temperatures.
It converges to a step function with the step formed at the
chemical potential $\mu$ when $T \rightarrow 0$.
The temperature dependent grand canonical density matrix 
is formally given by the operator relation
\begin{equation}
\rho(\beta ) = \Phi_{\rm FD}(H,\mu,\beta).
\end{equation}


The single-particle energy of a fermion system 
at a finite temperature is given by
\begin{equation}\label{ES}
E_s = Tr[H \rho(\beta )] 
= \sum_{i,j} \langle \phi_i|H|\phi_j\rangle 
\langle \phi_j|\rho(\beta)|\phi_i\rangle,
\end{equation} 
in some set of basis functions $\phi_i$. 
In this formulation the expression for 
$\rho(\beta )$ does not have to be calculated explicitly; instead
the Fermi-Dirac function can be expanded in Chebychev 
polynomials.
The Chebychev expansion technique is one of the most efficient ways of
approximating a function and the Chebychev functions $T_n(x)$ 
obey a simple two-step recurrence
formula, ($T_0 = 1$, $T_1 = x$),
\begin{equation}
T_{n+1}(x) = 2xT_{n}(x) - T_{n-1}(x).
\end{equation}
The products $\Phi_{\rm FD}(H,\mu,\beta)|\phi_i\rangle$ can be calculated 
efficiently using the two-step recurrence formula using only matrix-vector 
multiplications \cite{Goedecker94,Silver94,Wang94,Silver96,Baer97}. 
The Chebychev expansion technique has many advantages: for example, the costly 
matrix-matrix multiplications are avoided, and error accumulation
is small. However, compared to linear scaling purification techniques
it is fairly inefficient \cite{Daniels99}.  The problem is the slow linear
increase in polynomial order as a function of iterations in the two-step
recurrence formula. To improve the
efficiency, Liang {\it et al.} \cite{HG_ACS}
recently suggested an alternative approach, where the expansion
polynomials are not calculated by the two-step formula, but
by a direct expansion. This can be achieved with a computational 
complexity that scales with 
the square root of the polynomial order of the expansion ${\cal O}({\sqrt n})$.
This limit is optimal for a general polynomial \cite{Paterson73}.
However, by choosing the expansion with
a particular set of Pad\'{e} polynomials, we will show how 
the computational complexity can be reduced even further, 
scaling only linearly with the logarithm of the polynomial 
order ${\cal O}({\log n})$ or the inverse temperature ${\cal O}({\log \beta})$.


There are at least 19 different ways to calculate
matrix exponentials \cite{19ways}. Here we 
use one particular technique based on a Pad\'{e} approximation.
Consider the exponential function 
\begin{equation}
e^x = \left( e^{x/n}\right )^n =  
\left(\frac{e^{x/(2n)}}{e^{-x/(2n)}}\right)^n.
\end{equation}
A Taylor expansion to first order gives
\begin{equation}\label{Exp}
e^x = \lim_{n \rightarrow \infty} \left( \frac{2n + x}{2n - x} \right) ^n.
\end{equation}
This Pad\'{e} approximation can be used in
the Fermi-Dirac function and for the rescaled chemical potential
and inverse temperature \cite{Notation}, ${\mu'} = 1/2$ and $\beta' = 4n$,
\begin{equation}
\Phi_{\rm FD}(x,\frac{1}{2},4n) 
\approx \frac{(1-x)^n}{x^n + (1-x)^n}.
\end{equation}
At higher values of $n$ the approximation becomes increasingly better.
The choice ${\mu'} = 1/2$ is made to center the step of
the Fermi-Dirac function at $x=1/2$. In the interval $[0,1]$
the approximation is a continuously
decreasing function with a maximum of $1$ at $x=0$ and a minimum
of $0$ at $x=1$. This is the interval in which the
temperature dependent density matrix has its eigenvalues and
it is the interval where the Fermi-Dirac distribution is well
approximated already at fairly high temperatures.
It is therefore the interval around which we chose to perform 
the expansion. This choice requires an initial rescaling of
the Hamiltonian spectra around the interval $[0,1]$.
Let
\begin{equation}\label{G}
G_n(x) = \frac{x^n}{x^n + (1-x)^n}.
\end{equation}
then the Fermi-Dirac function 
for ${\mu' }= 1/2$ and $\beta' = 4n$ in the interval $[0,1]$ is approximated by
\begin{equation}\label{Pade}
\Phi_{\rm FD}(x,\frac{1}{2},4n) = \left[ e^{4n(x-1/2)} + 1\right]^{-1} \approx 1 - G_n(x).
\end{equation}
The polynomial order $n$ of the Pad\'{e} 
approximation is proportional to the inverse temperature since $ n = \beta'/4$. 
This means that the lower the temperature the better the approximation.
In practice, however, 
the approximation at the normalized energy interval $[0,1]$ 
is already very good at orders as low as $n \approx 5$.
An example
given in Fig.\ \ref{FD_G5}, which shows the Pad\'{e} approximation
$1-G_5(x)$, is virtually identical to the corresponding Fermi-Dirac
function with $\beta' = 20$ and ${\mu'} = 1/2$.  The inset shows the 
error. With the interval
$[0,1]$ equal to 1 Ry this example corresponds to a temperature
of 7894 K. At lower temperatures the approximation becomes
increasingly better.  The Pad\'{e} approximation in Eq.\ (\ref{Pade}) 
is only one alternative, but, as will be shown below, it turns out 
to be particularly simple and efficient.



A major advantage with the Pad\'{e} approximation in 
Eq.\ (\ref{Pade}) is how efficiently we can calculate 
high orders of $G_n$. The computational complexity is
very low thanks to the iterative relation
\begin{equation}\label{nested}
G_{k \times l} (x) = G_k(G_l(x)).
\end{equation}
In an operator expansion this corresponds to purifications, projecting
the eigenvalues towards 0 and 1.
In contrast to a more general polynomial expansion such as the
Chebychev expansion, which computationally scales at best with 
the square root of the polynomial order, ${\cal O}({\sqrt n})$  
\cite{HG_ACS,Paterson73}, or as ${\cal O}(n)$ if the two-step recurrence
formula is used, the iterative relation above makes it possible 
to reach the same order of expansion in only ${\cal O}(\log n )$ steps.
The same low logarithmic complexity is found generally in purification
expansion schemes that are based on iterative spectral projections.
The Pad\'{e} approximation of the Fermi-Dirac distribution can thus 
be used in a highly efficient purification scheme which calculates
the finite temperature density matrix $\rho(\beta )$ at a set 
of normalized inverse temperatures $\beta'=4n$. The purification 
algorithm can be described by
\begin{equation}\label{IP}
\begin{array}{ll}
X_1 &= F(H,\mu) \\
X_{i+1} &= G_m(X_i), ~~~ i = 1,2,\ldots , \log_m(n) \\
\rho(\beta )  & = I - X_{i+1}.
\end{array}
\end{equation}
The expansion order $n$ and thus the normalized inverse temperature $\beta'$
must be chosen so that the number of iterations $\log_m(n)$, is an integer.
The function 
\begin{equation}
F(H,\mu) = \alpha(H-\mu I) + 0.5I
\end{equation}
is a normalization function that rescales
all the eigenvalues of $H$ to the interval $[0,1]$,
with the chemical potential $\mu$ shifted to 
${\mu'} = 1/2$.
The chemical potential and spectral bound must thus be
known in advance.  The normalization factor 
\begin{equation}\label{rescale}
\alpha \approx \frac{1}{2} \min 
\left[(\mu-H_{\rm min})^{-1},(H_{\rm max}-\mu)^{-1}\right], 
\end{equation}
rescales the spectra and sets the temperature scale. 
The implicit purification scheme converges to the zero 
temperature density matrix for any value of $\alpha > 0$,
but the convergence is faster and the approximation 
is more accurate at higher temperatures if $\alpha$
is chosen to normalize the spectra around $[0,1]$.
The spectral bounds $H_{\rm max}$ and
$H_{\rm min}$ can be estimated by for example Lanczos'
algorithm or Gersgorin circles.
Generally we have that the temperature \cite{Notation}
$T = 1/(\alpha k_B 4n)$, 
where $n$ is the accumulated expansion order in Eq.\ (\ref{IP}).

Because of the rational form of $G_m(X_i)$ the purification scheme
is implicit. Assuming a finite orthogonal basis representation, 
a set of linear equations in Eq.\ (\ref{IP}) has to be solved in each 
step for $i = 1,2 \ldots,\log_m(n)$:
\begin{equation}\label{IP_eq}
\left[ X_i^m + (I-X_i)^m \right] X_{i+1} = X_i^m,
\end{equation}
which is given from the second step in Eq.\ (\ref{IP}) and from the 
definition of $G$ in Eq.\ (\ref{G}) along with its nested iterative
expansion property given in Eq.\ (\ref{nested}). 
Here we find another major advantage
with our particular choice of Pad\'{e} approximation. The left side 
system matrix $A_i = [ X_i^m + (I-X_i)^m]$ is symmetric and positive
definite for symmetric $X_i$'s with their spectra belonging to $[0,1]$. 
In fact, with increasing $i$, the system matrix $A_i$ converges to 
the identity matrix $I$.  The implicit equations are therefore very well 
suited for solutions with the linear conjugate gradient method 
\cite{CG_method}, that in turn, 
can efficiently exploit the close approximation of $X_i$ to the unknown 
columns of $X_{i+1}$, which becomes increasingly more accurate
and efficient towards the last iterations. 
Another possibly efficient alternative is the application of the sparse
approximate inverse (AINV) \cite{Benzi96,Challa99} that can be
expected to work well for this particular problem. However, this
approach has not been explored in the present study.


Alternative implicit purification schemes can also be derived from
various sign matrix expansions \cite{Kenney91,Beylkin99}. Sign matrix 
expansions are equivalent to purification. The only difference is that 
spectral projections are performed in the interval $[-1,1]$ instead of 
$[0,1]$, as in the case of purification. 


To analyze the efficiency of the algorithm compared with
explicit purification schemes,
we have chosen an $N\times N$ model Hamiltonian
with $N$ random diagonal elements. The overlap elements decay
exponentially as a function of
site separation on a randomly distorted simple cubic lattice.
This model represents a Hamiltonian of an insulator
that might occur, for example, with a Gaussian basis set in
density functional theory or in various tight-binding schemes.
The convergence is mainly determined by the occupation
and the band gap. The test Hamiltonian was therefore modified such that
all $N$ eigenvalues were uniformly distributed on $[0,1]$.
In this case the band gap $\Delta g = 1/N$, independently of 
the fractional occupation $f_{\rm occ} = N_e/N$. 
This simplifies the analysis and comparison
of the different methods, which otherwise are hard to perform
for a less idealized set of material systems.
After each iteration a numerical threshold $\tau = 1.0 \times 10^{-7}$
was applied and convergence was determined when the error
in energy $|E_{\rm approx} -E_{\rm exact}| <  1.0 \times 10^{-5}$ 
\cite{Notation}. The convergence criterion corresponds in practice to 
$T=0$. A comparison at $T \approx 0$ is necessary since 
the explicit purification schemes used in the comparison only give 
the zero temperature density matrix. At room temperature the computational 
effort with the implicit purification scheme is only slightly reduced 
because of the rapid convergence. 
The computational complexity was measured in number of matrix-matrix 
multiplications, where $N$ conjugate gradient steps, i.e. $N$
matrix-vector multiplications,  are counted as one
matrix-matrix multiplication.

Figure \ref{Comp_All} shows the computational cost for various
occupation factors.
The implicit purification scheme of order two (IP), 
i.e.\ with $m=2$ in Eq.\ (\ref{IP_eq}),
using $X_i$ as initial approximations to $X_{i+1}$,
is compared to the trace correcting scheme with second order
polynomials (TC2) by Niklasson \cite{Niklasson03},
the trace resetting asymmetric fourth order method (TRS4) by 
Niklasson {\it et al.} \cite{NiklassonSP4},
the grand canonical scheme with fourth order projections
(GC4) by Niklasson \cite{Niklasson03},
the grand canonical McWeeny purification 
scheme (McW) \cite{McWeeny60,Palser98}, 
and finally the canonical scheme (PM) by Palser and 
Manolopolous \cite{Palser98}.  The grand canonical 
schemes that require prior knowledge of the chemical
potential are indicated in the figure by bold italics. 


For small band gaps, i.e.\ high values of $N$, and with prior knowledge of $\mu$, 
the asymmetric GC4 method is the most efficient technique. 
The best performing schemes that require no prior knowledge of 
$\mu$ are the TC2 and TRS4 schemes.
The TC2 scheme is  more memory efficient since it only needs to calculate 
a second order polynomial in each iteration and intermediate storage 
needed in higher order expansions is avoided. 
However, it can not deal with degeneracy and fractional occupancy, 
which are addressed with the TRS4 scheme \cite{NiklassonSP4}.  
At low occupation the PM scheme becomes very inefficient. This sensitivity is
not seen for any of the other schemes.
The proposed implicit purification scheme is 
slower than the alternative explicit purification schemes except
for the PM scheme at low occupancies. However, it is the only method that,
for only a slightly increased computational cost,
correctly gives the temperature dependent Fermi-Dirac distribution of the 
single-particle eigenstates.  The implicit purification scheme
scales with the logarithm of the expansion order. This is an important
improvement over previous Fermi-Dirac operator expansion methods.
Whereas in general, a polynomial can be calculated with a computational
cost scaling at best with the square root of the order of the polynomial
\cite{Paterson73}, we restrict the polynomial approximation to a
nested form $f(f(\ldots f(x)\ldots ))$. In this case a high order
can be reached much more efficiently than for the general form.
This is the key idea behind purification expansions. 


In summary, we have proposed an implicit purification scheme for 
the calculation of the temperature dependent single-particle density 
matrix given as a Fermi-Dirac operator expansion in terms of
the Hamiltonian.  The method is useful in connection with linear 
scaling electronic structure theory and it has a
computational complexity that scales with  
the logarithm of the inverse temperature ${\cal O}(\log \beta)$ 
or as the logarithm of the polynomial expansion order ${\cal O}(\log n)$. 


Discussions with Matt Challacombe, Eric Chisolm, Siobhan Corish, 
S.\ Goedecker, C.\ J. Tymczak, and John Wills are gratefully acknowledged.

\begin{figure}
\caption {\small The Fermi-Dirac distribution (dashed line) compared to
the approximation $1-G_5(x)$ (circles) in Eq.\ (\ref{Pade}).
The inset shows the error, ${\rm Error} \times 10^3 = \bigl(\Phi_{\rm FD}(x,1/2,20)-
[1-G_5(x)] \bigr) \times 10^3 $.
\label{FD_G5}}
\end{figure}

\begin{figure}
\caption {\small Computational cost for various schemes at $T\approx0$. 
Grand canonical schemes requiring prior knowledge of $\mu$ are written
with bold italics. The N eigenvalues are uniformly distributed in $[0,1]$
and the band gaps are therefore $\Delta g = 1/N$ independent of the
fractional occupation $f_{\rm occ} = N_e/N$. 
\label{Comp_All}}
\end{figure}

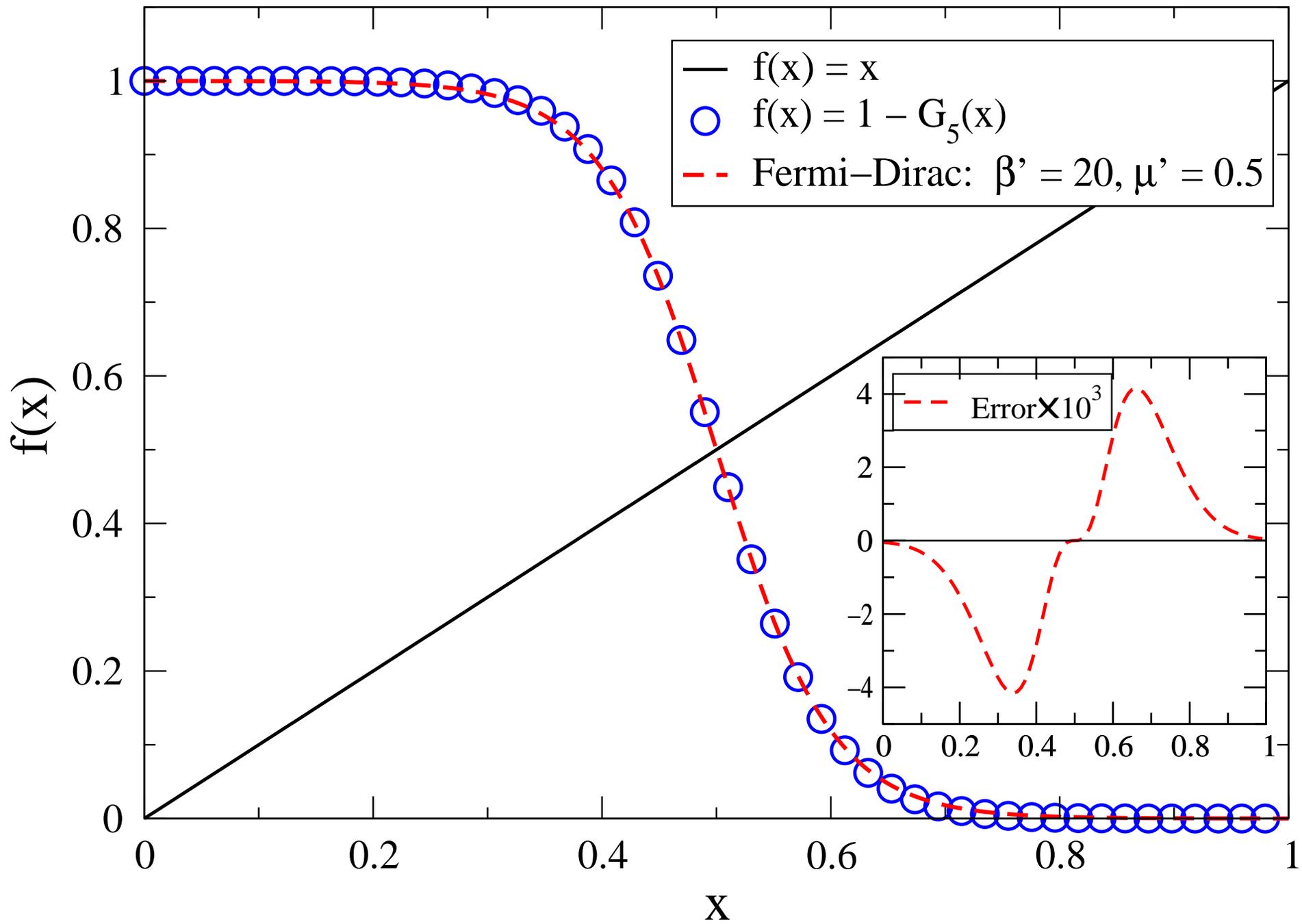
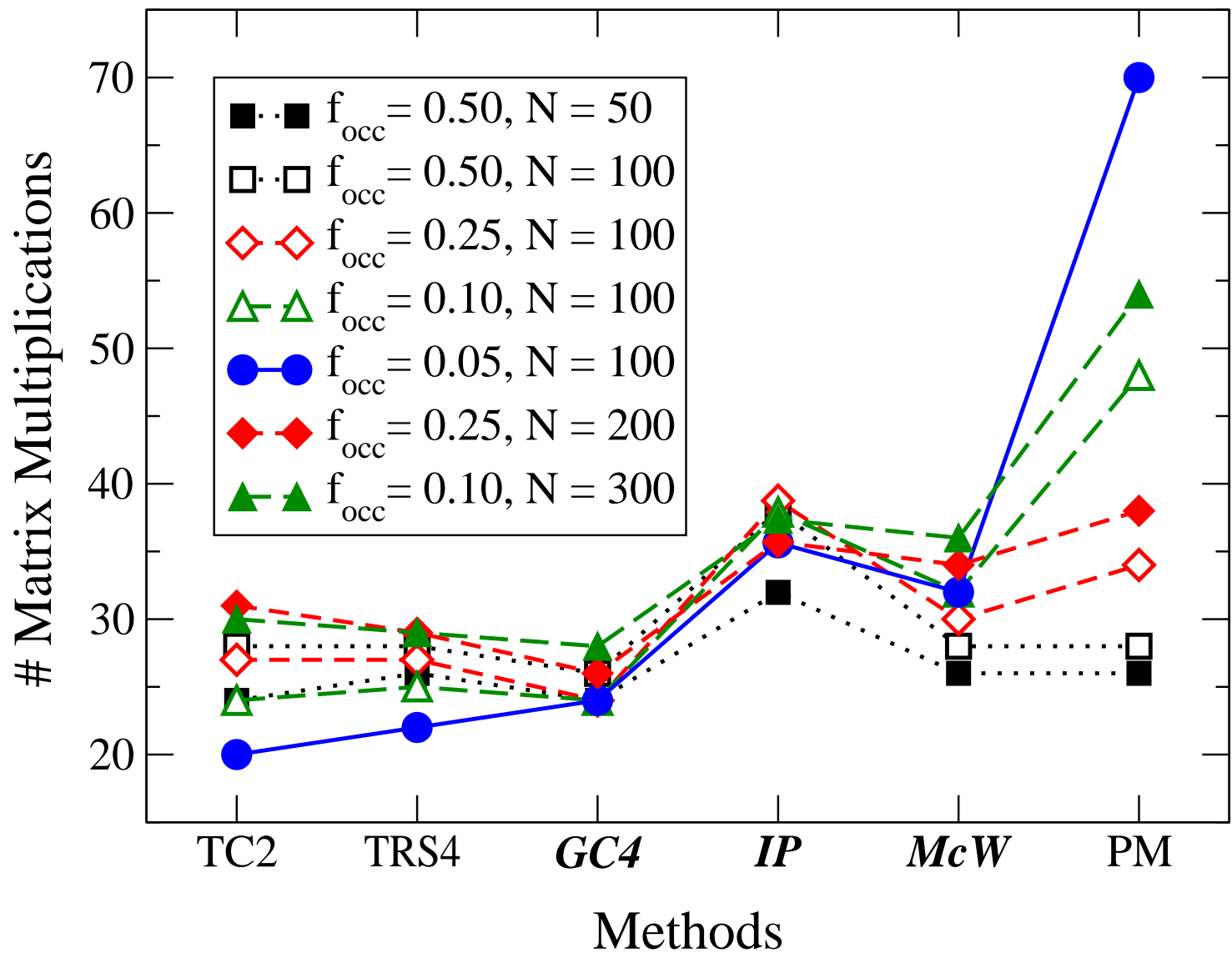
\end{document}